\documentclass[aps,prl,twocolumn,showpacs,preprintnumbers,amsmath,amssymb]{revtex4}
\usepackage{graphicx}
\usepackage{dcolumn}
\usepackage{bm}

\begin{document}

\title{Identifying neutrinos and antineutrinos in neutral-current scattering reactions}
\author{N.~Jachowicz, }
\email{natalie.jachowicz@UGent.be}
\author{K.~Vantournhout, J.~Ryckebusch, K.~Heyde}
\affiliation{Department of Subatomic and Radiation Physics,\\ Ghent University, \\Proeftuinstraat 86, \\ B-9000 Gent, Belgium.} 
\date{\today}

\pacs{25.30.Pt,13.88.+e,24.70.+s,26.50.+x}

\begin{abstract}
We study neutrino-induced nucleon knockout from nuclei. Expressions for the induced polarization are derived within the framework of the independent-nucleon model and the non-relativistic plane-wave approximation.  Large dissimilarities in  the nucleon polarization asymmetries are observed between neutrino- and antineutrino-induced processes.
These asymmetries represent a potential way to distinguish between  neutrinos and antineutrinos in neutral-current neutrino-scattering on nuclei.  We discuss astrophysical applications of these polarization asymmetries.
Our findings are illustrated for neutrino scattering on $^{16}$O and $^{208}$Pb.
\end{abstract}
\maketitle

In charged-current (CC) neutrino-scattering processes the determination of the nature of the incoming lepton is straightforward.  Conservation of lepton number implies that neutrinos generate an electron, muon or tau lepton, while inelastic scattering  of an antineutrino produces the corresponding antilepton. In either case, the charge of the outgoing lepton unambiguously reveals the nature of the incident particle.
In neutral-current (NC) weak processes however, the outgoing lepton is a neutrino, and the difference does not manifest itself in this clear way. 

Nonetheless, in a number of situations it can be very interesting to differentiate between neutrinos and antineutrinos. 
Speculations about CP-violation in neutrino oscillations involve interest in discriminating between neutrinos and antineutrinos to
find out whether they are indeed behaving in a different way \cite{boone,cp}.  However, CC reactions are not always accessible as neutrino-detection mechanism.   This is the case for $\nu_{\mu}$ and $\nu_{\tau}$ neutrinos over a rather wide energy range.

The difference between neutrinos and antineutrinos is of peculiar interest in supernova-neutrino physics.  Most of the energy released in a type II supernova explosion is radiated away by neutrinos  of all kinds. The dynamics of the supernova process is very sensitive to neutrino interactions.  
The fact that a neutron star is forming in the center of a core-collapse supernova favors neutrino  over antineutrino reactions.  
The neutrinos are escaping from close to the center of the star, carrying away information about the processes driving the explosion and the formation of a neutron star \cite{nnp,mat}.
As neutron matter is more opaque to neutrinos than to antineutrinos, they escape from different sites, with different energies.

Thus far, theoretical simulations have proven unable to generate a successful explosion, a problem possibly due to missing aspects in weak interaction physics \cite{nnp}.
An experimental study could shed light on this topical issue in stellar evolution. Confronting models with the observation of a supernova neutrino-flux would provide a stringent test for supernova models \cite{vogel,beacom1}. 
Neutrino-nucleus scattering is considered as a promising technique \cite{omnis,land} to detect supernova neutrinos.  These reactions allow to detect any kind of neutrino, and the nucleon emission  thresholds in nuclei are situated in an energy range well suited to extract information about the energy distribution of neutrinos. Studies of the folded cross sections show that a neutrino-nucleus based detector would be most sensitive to neutrinos with energies around 40 MeV \cite{ikke3}. At typical supernova energies,  $\nu_{\mu}$ and  $\nu_{\tau}$ neutrinos do not participate in CC reactions, so NC scattering constitutes an essential part of this study.
As a galactic supernova explosion is an extremely rare occurrence \cite{beacom1}, it is crucial to gather as much information as possible in the event.  Obtaining information about the time and energy distribution of the arriving neutrinos, and about their luminosities 
contributes to this objective. Distinguishing between neutrinos and antineutrinos can provide additional information.  

In recent years, a large number of theoretical studies on neutrino scattering from hadronic matter were carried out
including Fermi gas, random phase approximation, and shell model calculations \cite{ikke3,Ed,kim,kol1,albe,sing,ikke0,fuller,town,vol,ikke1,ikke2,volvol,volpenieuw}.  
Thereby, little or no attention has been paid to the polarization degrees-of-freedom for the ejected nucleons. 
In this letter, we demonstrate that in neutrino-nucleus processes the helicity-related differences between neutrinos and antineutrinos induce strong asymmetries in the polarization of the ejected nucleons.
Hereafter, an exploratory investigation of ejectile polarization in neutrino-induced nucleon knockout is conducted. Our derivations are done within the framework of the non-relativistic nuclear shell model, with a Woods-Saxon description for the bound nucleon states.   Expressions for the A($\nu$, $\nu'$ N) and A($\overline{\nu}$, $\overline{\nu}'$ N) cross sections are derived adopting a plane-wave description for the ejectiles.  Results of numerical calculations for NC $\nu$-induced nucleon knockout from $^{16}$O and $^{208}$Pb are presented for incoming neutrino energies from 10 MeV up to 500 MeV.

\begin{figure}
\vspace*{4cm}
\special{hscale=48 vscale=48 hsize=1500 vsize=300
         hoffset=0 voffset=0 angle=0 psfile="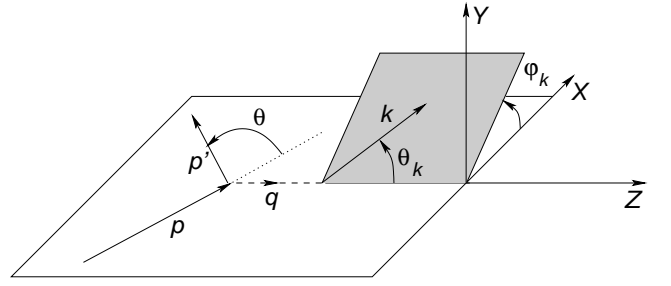"}
\caption{A neutrino with momentum $p$ is scattered from a nucleus over an angle $\theta$ thereby obtaining a final momentum $p'$.  The momentum transfer is $q_{\mu}=p_{\mu}-p_{\mu}'$.  The outgoing nucleon has momentum $k$, in a direction $\theta_k$ relative to the momentum transfer.  In the following, $\theta_{N}$ denotes the angle between the incoming lepton and the outgoing nucleon.}
\label{axis}
\end{figure}

Fig.~\ref{axis} displays the kinematic variables for the reactions 
we are considering.
As in NC neutrino scattering the outgoing lepton is not detected, 
the momentum exchange remains unknown and the missing momentum cannot be reconstructed.
Therefore, the direction of the outgoing nucleon $\theta_{N}$ is defined  relative to the direction of the incoming neutrino  and any ejectile's polarization study is bound to focus on the longitudinal spin components $s_N^l$ of the nucleon i.e. the spin component along the direction of its momentum.
The differential cross section can be written as
\begin{eqnarray}
{\frac{d^3\sigma}{d\Omega_{N}dE_{N}} =}(2\pi)^4  E_{N}k\;
\overline{\sum}_{f,i}
\left|\langle f\left|\widehat{H}_W\right|i\rangle\right|^2,\label{cs}
\end{eqnarray}
where $\widehat{H}_W$ represents the weak interaction Hamiltonian and the expression is averaged over all possible initial states, and summed over all available final states.  The transition matrix element  factorizes in a lepton and a hadron current, resulting in a 
lepton part that can be cast in the form
\begin{equation}\label{f4}
\frac{1}{2}\sum_{ss'}l_\mu l_\nu^*=\frac{p_\nu p'_\mu+p_\mu p'_\nu-pp'g_{\mu\nu}\pm ip^\alpha p'^\beta \epsilon_{\alpha\nu\beta\mu}}{(2\pi)^6 \varepsilon_i\varepsilon_f},
\end{equation}
with $p_{\mu}$ and $p_{\mu}'$ the incoming and outgoing lepton momentum, $\varepsilon_i$ and $\varepsilon_f$ the corresponding energies, and $\epsilon_{\mu\nu\rho\sigma}$ the four-dimensional Levi-Civita symbol.
Rewriting the hadron current $h^\mu=(h^0,\vec{h})$ in spherical components, 
the transitions are determined by the expression
\begin{eqnarray}
\lefteqn{\left|\langle\Psi_{A-1};k,\frac{1}{2}s^q_N;\,p'|\widehat{H}_W|\Psi_i;\,p\rangle
\right|^2=\frac{G_F^2}{2(2\pi)^6\varepsilon_i\varepsilon_f}}\nonumber\\
& &[p_0 h_0-\!\sum_{\lambda=\pm 1,z} (-1)^\lambda p_\lambda h_{-\lambda}][p'_0
  {h_0}^*-\!\sum_{\lambda=\pm 1,z} p'_\lambda{{h}_{\lambda}}^*]\nonumber\\
&+&[p'_0 h_0-\!\sum_{\lambda=\pm 1,z}(-1)^\lambda {p}'_\lambda h_{-\lambda}][p_0
  {h_0}^*-\!\sum_{\lambda=\pm 1,z} p_\lambda{{h}_{\lambda}}^*]\nonumber\\
  &-&[p_0p'_0-\vec{p}\cdot\vec{p}'][h_0h_0^*-\!\sum_{\lambda=\pm 1,z} h_\lambda{{h}_{-\lambda}}^*]\nonumber\\
&\pm&
\begin{vmatrix}
p_0 & p_+ & p_- & - p_z \\
p'_0 & p'_+ & p'_- & - p'_z \\
h_0 & h_+ & h_- & -h_z \\
{h_0}^* & -{h_-}^* & -{h_+}^* & -{h_z}^*
\end{vmatrix},\label{f7}
\end{eqnarray}
with $s_N^q$ the ejectile's spin component in the direction of momentum transfer.
The sign difference in the last terms of Eqs.~(\ref{f4}) and (\ref{f7})  arises from the spin projection operators for neutrinos and antineutrinos, the plus sign corresponding to $\nu$, the minus sign to $\overline{\nu}$ induced reactions.  After summing over the spins of the outgoing nucleon, Eq.~(\ref{f7}) reduces to an expression containing the well-known vector and axialvector Coulomb, longitudinal, and transverse electric and magnetic cross section contributions.
When the  cross section  is not averaged over the outgoing nucleon polarizations, various interference terms survive. 
The dominating parity violating axial contributions   in neutrino scattering make the transverse-transverse interference terms prominent
\cite{prep}.

The differences between neutrino- and antineutrino-induced ejectile polarizations can now be inferred from a number of observations.
First, the cross sections get their largest contribution from the transverse response \cite{ikke0}. Hence, the way the differences between neutrinos and antineutrinos affect the spin of the outgoing nucleon stems mainly from the spin properties of the transverse part of $\vec{h}$. The transverse response is dominated by the terms $h_+{h_+}^*$ and $h_-{h_-}^*$ of Eq.~\eqref{f7}.  The term $h_+{h_+}^*$ provides the prevailing contribution to the cross section for interactions with an ejectile spin aligned with the momentum transfer, and the term
$h_-{h_-}^*$ is prominent for outgoing nucleons with their spin antiparallel to the momentum transfer \cite{prep}. Second, the transverse dominance induces  a  preference for backward scattering of the neutrinos \cite{ikke1,ikke2}. Therefore, in the majority of scattering reactions the momentum exchange $\vec{q}$ between neutrino and nucleon, and the momentum of the incoming neutrino are aligned.  
This results in an outspoken preference for forward ($\theta_{N}$=0) emission of the nucleon. Hence, the direction of the outgoing nucleon tends to be aligned with the momentum exchange and the $h_+{h_+}^*$ and $h_-{h_-}^*$ contributions directly affect the longitudinal components $s_N^l$ of the ejectile's spin.
The transverse contribution is dominated by
\begin{eqnarray}
\lefteqn {(l_-{l_-}^* h_+{h_+}^* + l_+{l_+}^* h_-{h_-}^*)_{\stackrel{\nu}{\overline{\nu}}}} \nonumber\\
&&=S(h_+{h_+}^*+ h_-{h_-}^*) \mp A(h_+{h_+}^* - h_-{h_-}^*)\label{p1}\\
&&=(S\mp A) h_+{h_+}^* + (S\pm A) h_-{h_-}^*, \label{p2}
\end{eqnarray}
with the symmetric and antisymmetric kinematic factors
\begin{subequations}
\begin{align}
\begin{split}
S =&
\frac{2\sin^2\frac{\theta}{2}(\epsilon_i^2+\epsilon_f^2+2\sin^2\frac{\theta}{2}\epsilon_i\epsilon_f)}{(2\pi)^6(\epsilon_f^2+\epsilon_i^2-2\epsilon_i\epsilon_f\cos\theta)}, \label{s}\end{split}\\\begin{split}
A =&
\frac{2\sin^2\frac{\theta}{2}(\epsilon_i+\epsilon_f)}{(2\pi)^6\sqrt{\epsilon_f^2+\epsilon_i^2-2\epsilon_i\epsilon_f\cos\theta}}.\label{a}							  \end{split}
\end{align}
\end{subequations}
Eq.~(\ref{p1}) establishes the differences between $\nu$ and $\overline{\nu}$ cross sections : the transverse term consists of a part that is the same for neutrinos and antineutrinos and an antisymmetric contribution, with opposite signs for the lefthanded neutrinos and righthanded antineutrinos. 
After a reordering of  these contributions 
another aspect of the helicity related differences between neutrinos and antineutrinos shows up in Eq.~(\ref{p2}).
\begin{figure}
\vspace*{6.5cm}
\special{hscale=35 vscale=35 hsize=1500 vsize=300
         hoffset=-17 voffset=200 angle=-90 psfile="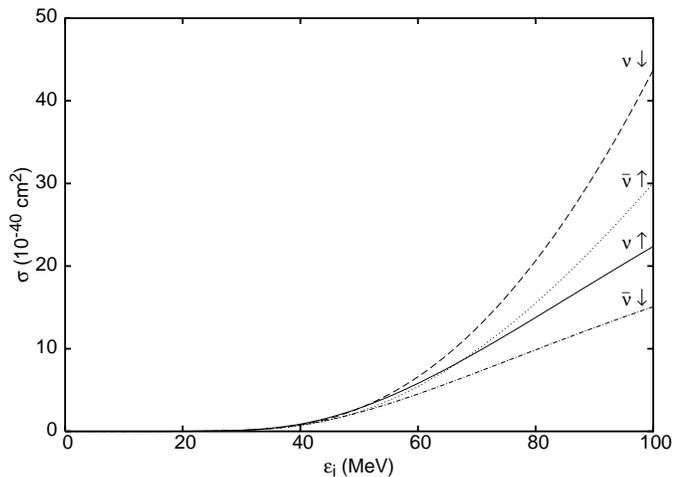"}
\caption{Longitudinal spin-up and spin-down proton knockout cross sections for NC neutrino- and antineutrino-scattering on $^{208}$Pb, as a function of the incoming lepton energy.}
\label{updown}
\end{figure}
From the expressions (\ref{s}) and (\ref{a}), it emerges that the kinematic factors
$S$ and $A$ are of the same order.
As a consequence, for neutrinos the forfactor $S-A$ of the 'spin up' contributions in Eq.~(\ref{p1})
becomes very small, resulting in a suppression of $s_N^l=\uparrow$ nucleon knockout,  while the $h_-{h_-}^*$ contribution is enhanced by the constructive effect of the factor $S+A$.  The $\theta$ dependence of Eqs.~(\ref{s},\ref{a})  shows that this effect is largest for backward lepton scattering,  
thus acting coherently with the aforementioned preferential scattering direction and enforcing the polarization effect.
For antineutrinos the asymmetry is completely reversed.

\begin{figure}
\vspace*{6.5cm}
\special{hscale=35 vscale=35 hsize=1500 vsize=600
         hoffset=-16 voffset=200 angle=-90 psfile="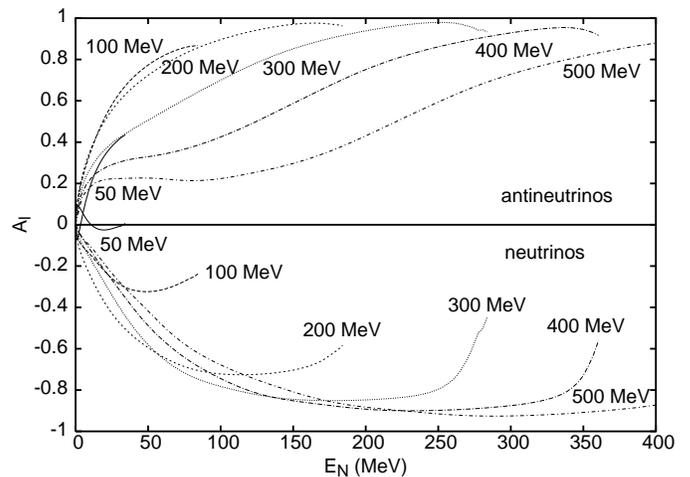"}
\caption{ Longitudinal asymmetry A$_{l}$ as a function of the ejectile energy for different incoming (anti)neutrino energies on a $^{16}$O target.}
\label{asymfig}
\end{figure}
Both aspects are illustrated in Fig.~\ref{updown}. The transverse interference terms are generally negative, resulting in   $\nu$ cross sections being larger than their $\overline{\nu}$ counterparts. The $\nu$ cross section is dominated by $s_N^l=\downarrow$ nucleons, the $\overline{\nu}$ cross section is characterized by a prominence of $s_N^l=\uparrow$ nucleons.  
  The increasing tendency for forward nucleon knockout with rising incoming neutrino energy causes the different energy dependence of the cross sections in Fig.~\ref{updown}.  With increasing incoming neutrino energies, the forward contribution becomes larger, $s_N^q$ and $s_N^l$ tend to be aligned, and the longitudinal spin behavior  of the reaction becomes even more closely related to this  predicted by Eq.~(\ref{p2}), while the other polarizations are suppressed. Hence $s_N^l=\downarrow$ neutrino and $s_N^l=\uparrow$ antineutrino cross sections gain in importance.
\begin{figure*}
\vspace*{6.5cm}
\special{hscale=35 vscale=35 hsize=1500 vsize=600
         hoffset=-16 voffset=200 angle=-90 psfile="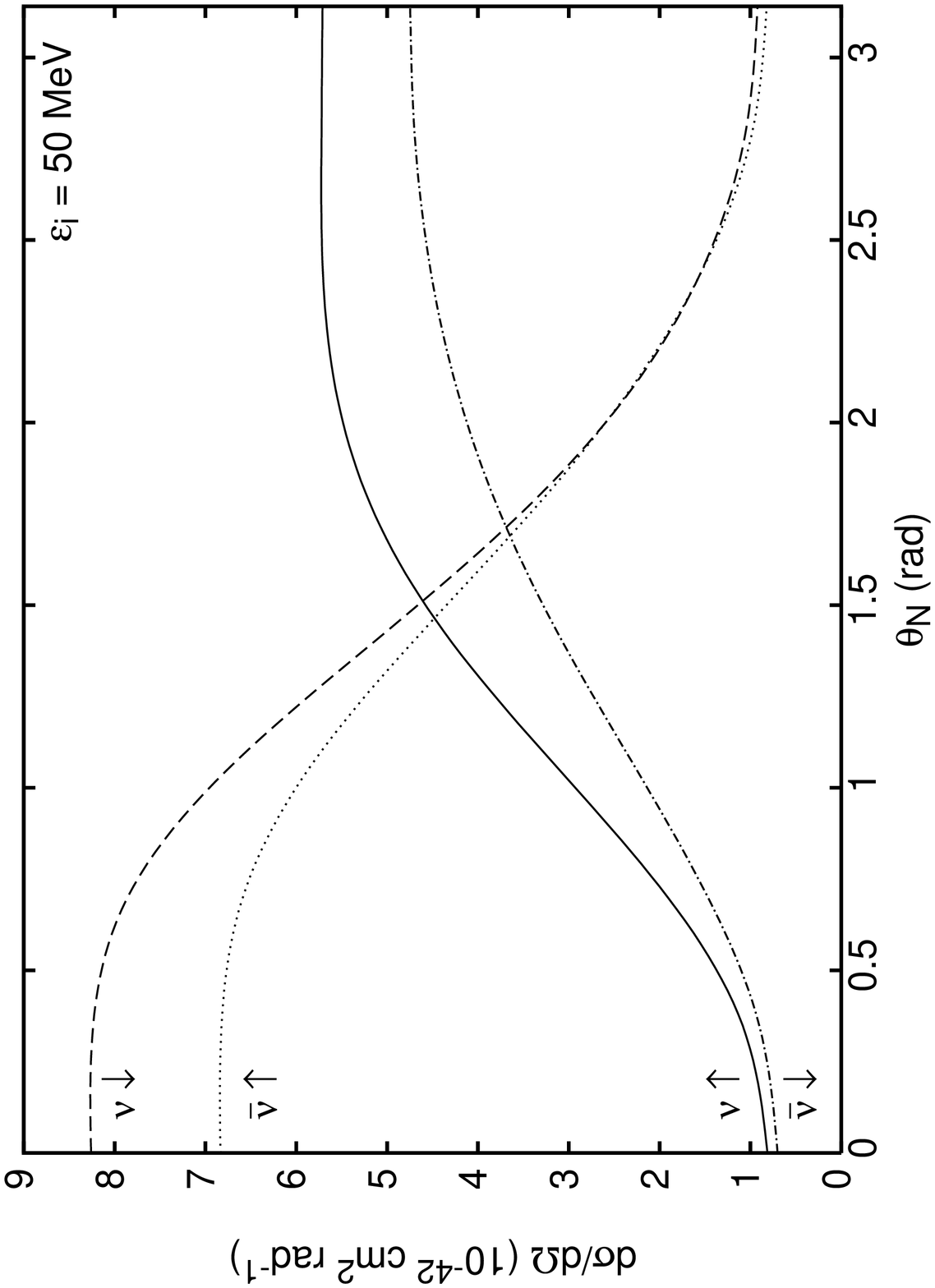"}
\special{hscale=35 vscale=35 hsize=1500 vsize=600
         hoffset=240 voffset=200 angle=-90 psfile="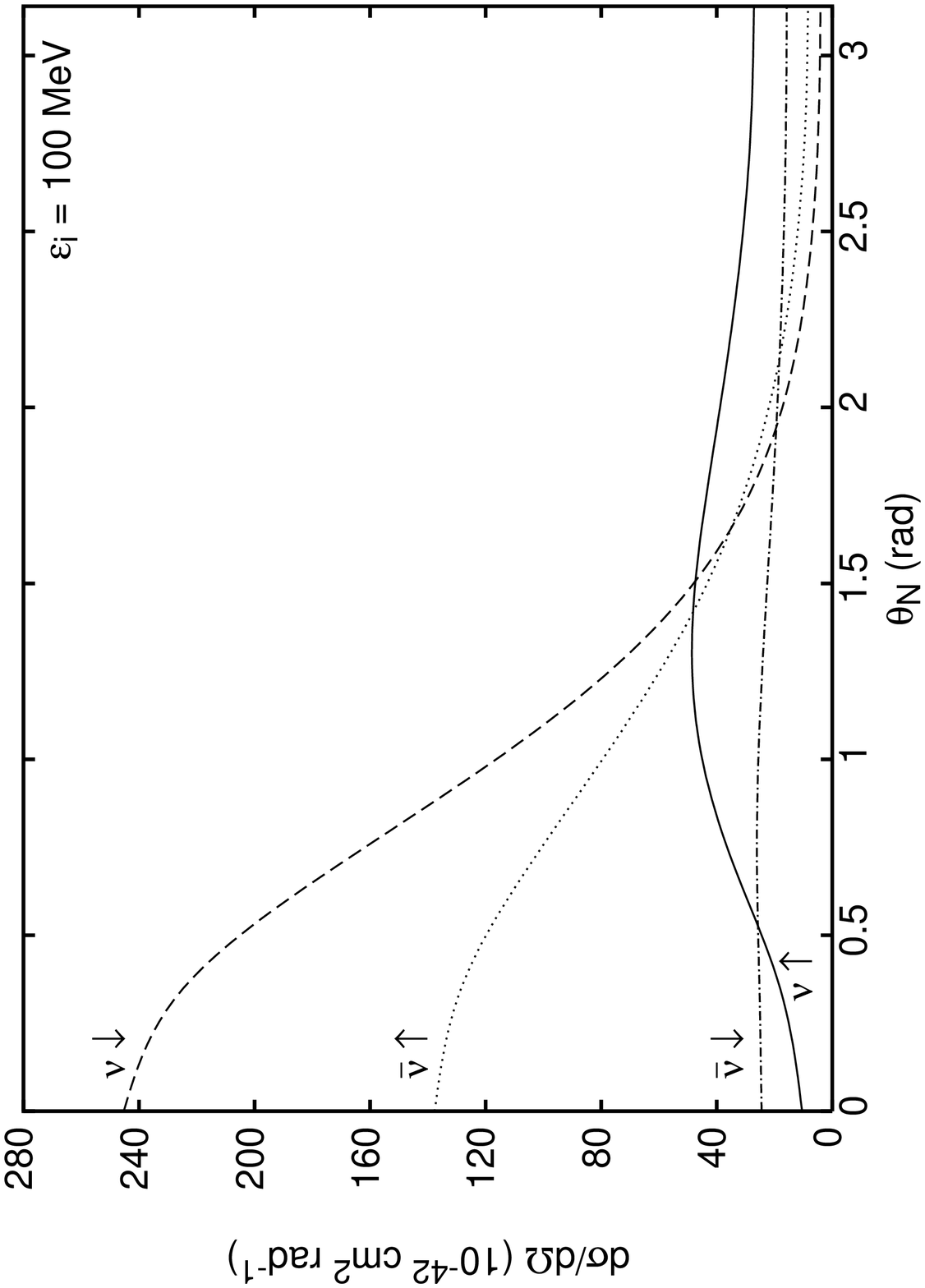"}
\caption{Comparison between $s_N^l=\uparrow$ and $s_N^l=\downarrow$ nucleon knockout for neutrino and antineutrino scattering off $^{16}$O as a function of the scattering direction of the outgoing nucleon $\theta_{N}$. The  left panel presents results for incoming lepton energy $\varepsilon_i=50$ MeV, the right panel corresponds to $\varepsilon_i=100$ MeV.  }
\label{O16}
\end{figure*}

Fig.~\ref{asymfig} investigates the longitudinal polarization asymmetry for neutrinos and antineutrinos
\begin{equation}
A_l=\frac{\sigma(s_N^l=\uparrow)-\sigma(s_N^l=\downarrow)}{\sigma(s_N^l=\uparrow)+\sigma(s_N^l=\downarrow)},
\end{equation}
over a large energy range.  The
polarization differences between neutrinos and antineutrinos result in $A_l$'s which are very large and of opposite sign.  The asymmetries $A_l$  approach their maximum value  with growing neutrino energies.  The effect is most pronounced when the major fraction of the energy exchange is  transfered to the outgoing nucleon.  

The noted dissimilarity in $A_l$  becomes even more obvious when examining the cross section as a function of the outgoing nucleon's scattering direction.  Fig.~\ref{O16} shows very large asymmetries between neutrinos and antineutrinos in the spins of the outgoing nucleon for $\theta_{N}$=0. 
The observation of a  forward nucleon with spin antiparallel to its momentum, makes it roughly a factor 15 more probable that a neutrino induced the reaction.
This effect remains large over an angular range of approximately 60$^{\circ}$.
For larger scattering angles, the effect of the additional SU(2) rotation needed to align the spin of the outgoing nucleon with its momentum becomes important. For the suppressed backward scattering, the 180$^{\circ}$ SU(2) rotation completely reverses the asymmetry effect.  Our calculations predict the asymmetry in the angular cross section to be  very large at incoming neutrino energies as low as 25 MeV.  The size of the observed asymmetries, the obvious opposite behavior of  $\nu$ and $\overline{\nu}$  induced processes, and the prominence of forward scattering
is such that correlations and final-state interactions (FSI) cannot be expected to distort the whole picture.  Drawing on conclusions obtained in electron-scattering from nuclei, the impact of secondary effects like FSI always tends to cancel when addressing ratios like the quantity $A_l$ \cite{janendimi}.

From an experimental point of view, our suggestions benefit from the fact that the crucial information is revealed by the asymmetry's  sign, and there is no need to measure absolute cross sections.  The asymmetries are large, particularly for forward nucleon knockout where most strength resides.  Moreover, incoming and outgoing neutrino beams are fully polarized by the weak interaction. 

In summary, we derived neutral current neutrino and antineutrino scattering cross sections, investigating the induced polarization of the outgoing nucleon.  We observe large asymmetries, with an opposite sign for neutrino- and antineutrino-induced processes.
Neutrinos preferentially eject nucleons with their spins anti-parallel to their momentum, the opposite behavior is noted for  antineutrinos.
The effect is most pronounced  considering the angular cross sections.
We propose the polarization of the ejected nucleons as an efficient way to discriminate between neutrinos and antineutrinos in neutral current  neutrino                                scattering off nuclei.
\acknowledgments
 The authors are grateful to the Fund for Scientific Research (FWO) Flanders and to the University Research Board (BOF) for financial support. They would like to thank D.~Ryckbosch for interesting discussions.

\end{document}